# On-sky results of the adaptive optics MACAO for the new IR-spectrograph CRIRES at VLT

J. Paufique[1], P. Biereichel, B. Delabre, R. Donaldson, R. Esteves, E. Fedrigo, P. Gigan[∀], D. Gojak, N. Hubin, M. Kasper, U. Käufl, JL. Lizon, E. Marchetti, S. Oberti, JF. Pirard, E. Pozna, J. Santos, S. Stroebele, S. Tordo

European Southern Observatory (ESO), Karl-Schwarzschildst. 2, 85748 Garching, Germany
[∀] LESIA, Observatoire de Paris Meudon, 5 place Jules Janssen, 92195 Meudon, France

## 1. ABSTRACT

The adaptive optics MACAO has been implemented in 6 focii of the VLT observatory, in three different flavors. We present in this paper the results obtained during the commissioning of the last of these units, MACAO-CRIRES. CRIRES is a high-resolution spectrograph, which efficiency will be improved by a factor two at least for point-sources observations with a NGS brighter than R=15. During the commissioning, Strehl exceeding 60% have been observed with fair seeing conditions, and a general description of the performance of this curvature adaptive optics system is done.

## 2. MACAO-CRIRES: high spatial resolution for high spectral resolution, and its specificities

CRIRES is a first-generation instrument of the VLT; installed at the Nasmyth focus of ANTU, it is providing high-resolution spectra in the NIR ($R=10^5$ in the 1-5 micron range). Harmoniously completing the VLT instrumentation as can be seen in Figure 1, it is a unique instrument offering such a resolution on 8-m class telescopes. High-resolution is up to now limited to the visible[i], 1-5 micron spectroscopy limited to roughly $10^{4}$[ii], and although some other experiment allow to explore such high-resolution in the infrared[iii], their use is limited to longer wavelength, and to smaller telescope diameters. Therefore, CRIRES, by allowing resolving the H-lines and providing high spatial resolution, will allow studying deeply embedded stars as well as close binaries, providing new insights on the kinematics of these stars; it will as well allow unprecedented studies of solar-system objects[iv].

CRIRES uses in normal operation a 0.2" slit size on the sky; the AO of CRIRES will provide a corrected beam for natural guide-star as dim as $M_R$=16. The concept used for the AO of CRIRES is mostly based on a series of curvature AO-systems, developed for the VLTI and for SINFONI[v,vi], already in routine operation in the Paranal Observatory.

CRIRES has started its commissioning process. While the –cold– spectrograph is in reintegration in the Paranal Observatory by the time of writing this paper, the AO part, MACAO-CRIRES, has already produced its first light. In the present work we present MACAO-CRIRES principle and performances and the first results obtained on-sky with our system.

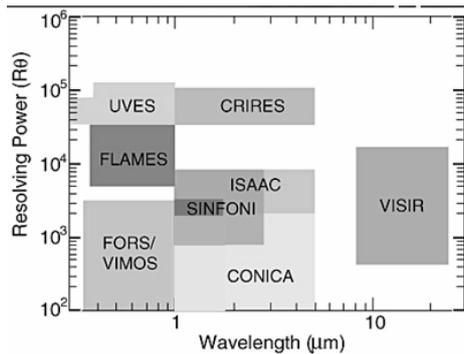

**Figure 1:** spectral resolution and wavelength coverage for VLT instruments. CRIRES extends the high spectral-resolution capability of the VLT towards infrared

---

[1] Contact: jpaufiqu@eso.org

## 3. Requirements of high spectral resolution

A spectrograph resolution is given by:

$$R = \frac{W(\sin\alpha + \sin\beta)}{\Phi D}$$

where $R$ is the resolution, $\sin\alpha\ resp.\sin\beta$ are respectively the incidence and diffraction angles, $W$ is the length of the beam on the grating, $\Phi$ is the slit width and $D$ is the diameter of the telescope.

The parameter $W$ influences directly the size of the major optical components (grating and spectrograph main optics), which is why the slit size has to be reduced so as to keep the instrument size (and weight) within acceptable limits. A slit size of 0.2" has been chosen so as to trade-off between the throughput of the spectrograph and the optics size.
In this frame, the AO brings a key advantage by roughly doubling the throughput of the instrument in the slit.

## 4. The MACAO projects

The MACAO project is based on two major lines:
- A standardization of the wavefront sensor components, common to 6 units,
- The use of the proven technique of curvature sensing

The standardization of the components allows the parallel development of three different instruments while minimizing the separated development to the essential instrument-related specifics (field-selector, piston issues mostly).
The curvature technique has proven to be an optimal technique for reaching diffraction-limited operation with a minimal number of actuators. In the MACAO systems, 60 actuators are used on the DM (deformable mirror), providing Strehl ratios between 50% and 60% for bright stars and typical Paranal seeings. This type of wavefront sensing is less linear than other techniques like Shack-Hartmann, but still provides key advantages, like the possibility to use only one photometric channel per subaperture, which drove us towards the use of virtually 0-noise APD sensors (Avalanche Photodiodes), following the combination bimorph-mirror/APD-curvature sensing introduced by Roddier for the CFHT adaptive optics PUEO.
Among the MACAO systems, two families exist:
the MACAO-VLTI family, based on the use of a 100 mm diameter bimorph mirror and a translation stage located in a telecentric beam, translating the whole wavefront sensor unit to center the guide-star, and
the CRIRES-SINFONI family, using a 60 mm diameter bimorph mirror, and a field selector located in a telecentric beam, keeping the wavefront sensor unit steady, at the price of a beam incidence variable within the field. The beam incidence being variable, it is therefore required to have a coordination between the field selector position and the tip-tilt of the membrane mirror (done through a gimbal mount on which the membrane mirror is mounted).

### 4.1. MACAO-CRIRES general description

MACAO-CRIRES is based on this second principle, and a top view of the specific part of MACAO-CRIRES is shown in Figure 2. The alignment of the different parts both in star-imaging and pupil-imaging is of course critical, as we need to measure the phase in a localized area, so as to be able to compensate for it. The beam, focusing right after the field lens, is collimated by the field selector lens, and then focused onto the membrane mirror by the imaging lens.
Concerning the pupil imaging, the pupil is sent at infinity by the field lens, then reimaged by the field selector lens at the level of the focus of the imaging lens, which send it therefore at infinity in the membrane mirror space.

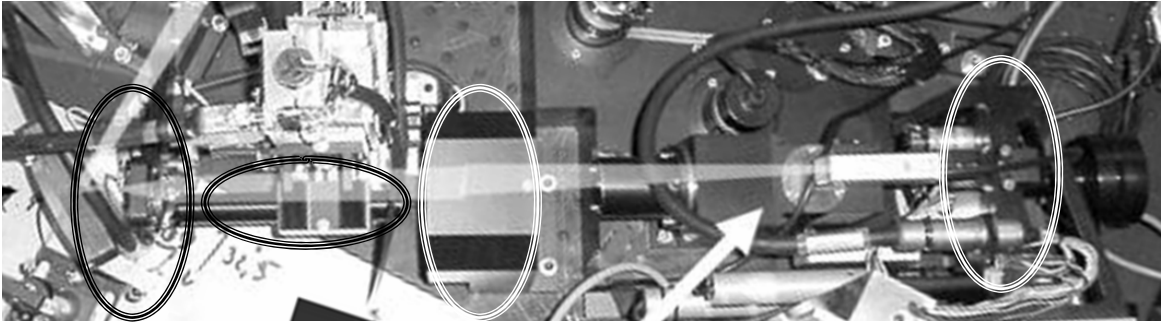

**Figure 2: close-up view of WFS are; from left to right, the field lens setting the entrance beam telecentric, the field selector lens, mounted on the field selector stage, the imaging lens, covering the whole field of view of CRIRES, and the membrane mirror, mounted in its gimbal mount. The arrow points to the wavefront sensor box, forming the intra- and extra-focal images of the beam on the lenslet array (below the support table, not visible on this picture)**

The wavefront sensing can then start. The vibration of the membrane mirror creates two alternate positions of the image of the pupil, allowing a comparative measurement of the flux in- and out-of-focus. The contrast between both measurements allows measuring the curvature error at the level of the pupil. A control matrix is then applied with an integral gain to compensate with the DM the aberrations measured. The principle of the different controls present in MACAO-CRIRES is summarized on Figure 3.

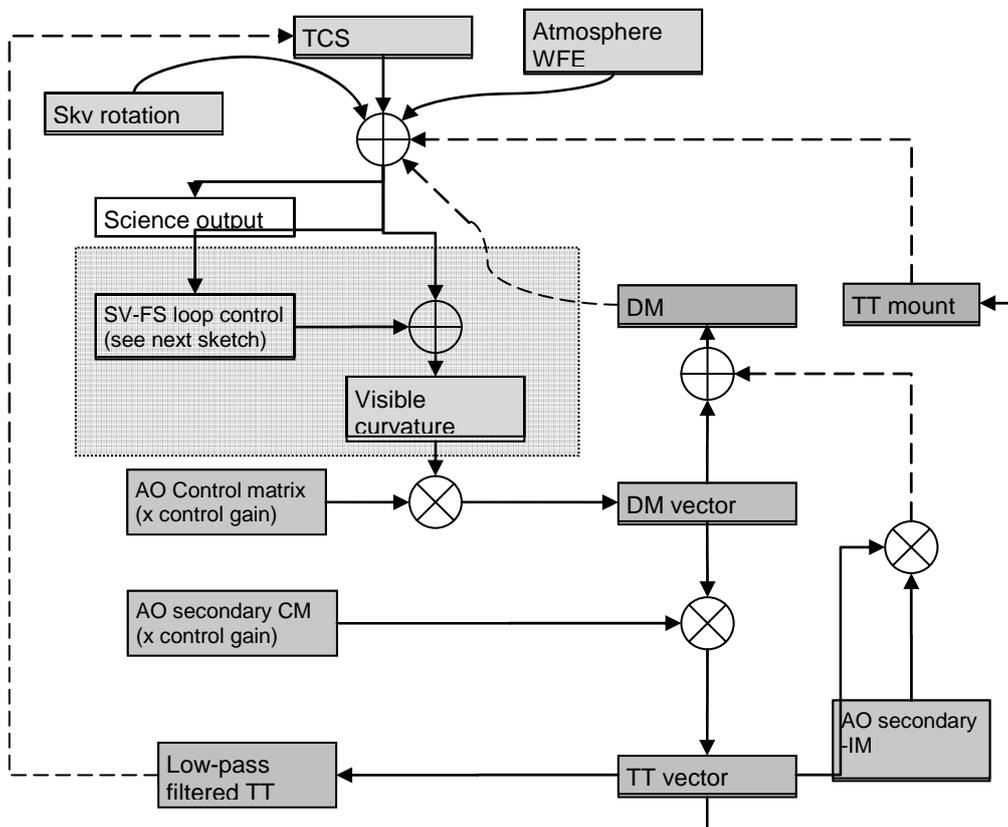

**Figure 3: MACAO-CRIRES loops. The TCS is the telescope control software, managing the whole telescope controls (pointing, active optics, ...). The SV-FS loop stands for the control scheme of the relationship between slit viewer images and field selector control, described in a further section**

Their application requires a number of calibrations to be done, so as to ensure the correction quality and reliability.

## 4.2. MACAO-CRIRES calibrations

MACAO-CRIRES requires 8 main calibrations to be able to operate in full:
The RoC (radius of curvature) of the membrane mirror must be calibrated with respect to the voltage applied to the membrane,
The phase lag between the membrane excitation voltage and the optical actuation on the beam must be calibrated to determine the sign of the curvature and to maximize the signal received when vibrating the membrane
interaction matrices for different RoC must be measured, so as to allow the computation of an optimal control matrix when a RoC is selected to optimize the correction with respect to the seeing and target extension,
a flat reference voltage vector to apply to the DM must be measured from time to time, to allow for an optimal open-loop wavefront quality (non-AO observations, calibrations, preset of the observations).
a secondary interaction matrix is measured between the deformable mirror and the tip-tilt mount, so as to relieve the low-bandwidth component of the tip-tilt correction vector requested to the AO-system,
a tip-tilt offload command must be as well calibrated so as to send appropriate telescope offset to correct for tip-tilt mount biases or slow drifts (error on proper motion estimation, on target coordinates or on differential refraction, for example)
the coordination between the field selector position in the field and the gimbal mount position, and
a field selector focus versus field position to correct for the combine field curvature of the instrument and of the telescope.
Most of those calibrations have been described all along the development of the MACAO project, we will therefore only summarize here the result of the optimizations done on the key aspects:

### 4.2.1. Radius of Curvature measurement

The measurement of the RoC is done in the MACAO-systems on a calibration source, through the application of a focus term to the field selector, which in return produces a strong focus term on the phase measured by the WFS, depending on the optical gain produced by the membrane mirror, ie inversely proportional to the RoC applied.

### 4.2.2. phase-lag measurement

For calibrating the phase-lag, a given voltage is applied to the DM, while the lag between membrane actuator (a simple voice-coil) and the optical signal received (the curvature) is monitored. The lag is tuned by dichotomy so that the minimum curvature is reached; this value is exactly located at 90° from the maximum signal given by the WFS.

### 4.2.3. interaction matrices

The IMs (interaction matrices) are at the very heart of the control of an AO system. Their inversion allows estimating the optimal voltage to apply to the DM to compensate for the WFE observed by the WFS. In our MACAO systems, the conditioning of the interaction matrices was a key issue, which is well described by Oberti et al. (2002)[vii]. All MACAO systems are now using a system mode approach, optimal in the case of an intrinsically non-linear WFS like curvature sensors. The IM acquisition is done at a rather high frequency, so as to freeze the turbulence in the instrument, which brings a key advantage in the case of MACAO-VLTI, and a more marginal one for SINFONI and CRIRES, as these have much smaller instrumental path length (by a factor 10) and maximum beam diameter (by a factor 5). Nevertheless, the homogeneity of the procedures gives way to a easier maintenance of the devices between all MACAO systems.

After a first blind iteration, the IM is computed recursively by applying a system-mode IM measurement.

This method consists in applying a set of 60 voltage vector to the DM, each of these vector expressing one of the eigenmodes of the WFS. This way, the intensity of the modes can be increased at the maximum value, so as to stay in the linear range of both the DM and the WFS. This way, the signal on each mode is maximised, and allowed us reaching a highly improved performance on our systems. A typical IM is given as an example in Figure 4, where the weak signal visible on the neighbouring subapertures is a sign of the quality of the alignment of the WFS and of the linear behaviour of the WFS.

An interaction matrix is acquired for the whole usable range of RoC, allowing an optimization of the performance as well for extended sources, and seeing characteristics.

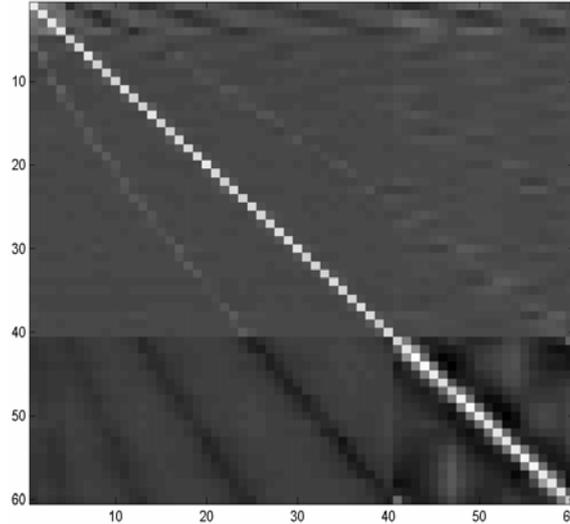

**Figure 4: typical interaction matrix given by MACAO-CRIRES. The curvature signal is given as a function of the electrode number to which a given voltage is applied. Note that the curvature sensor is very localized, due to the matching keystone-geometry of the electrodes/subapertures, with the exception of the outer electrodes, located outside of the pupil, and the outer sensors, sensitive mostly to the local tip-tilt of the beam, hence the darker area in the 20 lower rows, corresponding to the outer subapertures.**

### 4.2.4. Secondary interaction matrix

The secondary interaction matrix (and its inverse, the secondary control matrix) are used to compute the tip-tilt vector to offload on the tip-tilt mount. Its computation is done by applying a small tip (resp. tilt) of typically 4% of the maximum stroke -0.2 arcsec- to the TTM (tip-tilt mount), and closing the loop, while measuring the DM offset required to zero the curvature on the WFS. The matrix with the two DM-tip and DM-tilt vectors is then inverted and normalized so as to get the secondary control matrix.

### 4.2.5. tip-tilt offloading

The TTM has a wide and fast response, but its range is limited to +/-5", and a tip-tilt offload is used in all MACAO-systems, so as to avoid slow drifts to saturate the correction, ultimately causing an loop error. The procedure for the offload can be practically the same as for the secondary interaction matrix. Nevertheless, the offsetting being done at a very low-bandwidth, we rely on the North calibration of the derotator of CRIRES to offload in RA and DEC the telescope.

### 4.2.6. field selector calibrations

The FS (field selector) is a key component of CRIRES, as it is involved in both the AO system and the slit guiding. A calibration similar to the one performed for SINFONI is done, to coordinate the tip-tilt of the GM (gimbal mount –of the membrane mirror-) with the FS-unit position in the field. A combination of the movements of the FS and of the calibration carriage (holding the source, and allowing for only one translation in addition to the focus) provides a fair coverage of the field of view. The error with respect to the existing table is then recorded, and used to define refined parameters for the calibration of the coordination table. The model is a $2^{nd}$ degree polynomial function of the X and Y coordinates of the FS, providing a final accuracy better than 10% of a subaperture. A typical plot of residual errors is shown in Figure 5, corresponding to the one of first iteration done in the laboratory, over a limited field of view.

A similar calibration between the field selector X-Y and Z axis is done, so as to remove the telescope and instrument field-curvature when guiding off-axis. This off-axis error is usually overwhelmed by anisoplanetism errors, hence not critical for the performance of the system.

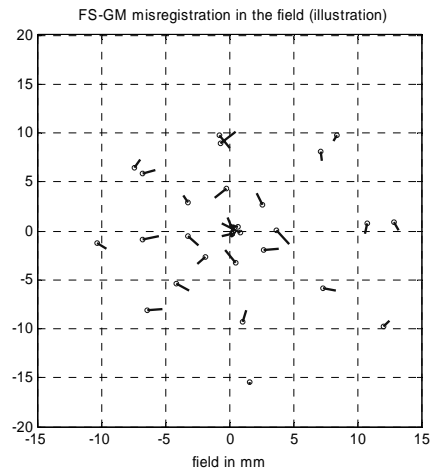

**Figure 5: FS-GM calibration error graph. The geometry of the points is determined by the combination of the derotator position and a lateral movement of the source. The lines indicate the error in fraction of sub-pupil; systematic errors can be accounted for, while the limit of the calibration lies within a few percent of a subaperture.**

**4.2.7. field-selector slit-viewer calibration**

The WFS is using a visible star to close the loop. Therefore, misestimating the wavelength of guiding, or the atmospheric refraction creates a slow drift of the star position on the slit. As the CRIRES spectrograph isn't sampling the diffraction spot, and barely the slit size (2 pixel), it is difficult to disentangle a slit centering and an absorption line dissymmetry. Therefore, a slit viewer has been implemented, to allow guiding at the observing wavelength (J to K-band) and optimize as well the throughput of the system. The principle of the control is presented in Figure 6. The centroid is computed for a small window (typically a 40 pixel box around the slit center). The centering error value is sent to the field selector, which will move in the opposite direction, so as to center the star image on the slit. The differential refraction model of the VLT observatory is used in feed-forward so as to decrease the requirements on the controller.

Note that the AO provides a diffraction-limited spot, so that typically 30% to 60% of the light is focused within a disk of 80 mas of diameter. The slit being 0.2" wide, this means that for centering errors below +/-60 mas, the centering measurement will be non-linear, and dependent on the atmospheric conditions. Though we observed in the laboratory the source with a turbulence generator, it is likely that lots of fine-tuning will be required on-sky to make the operation reliable and performing.

In spite of the non-linearity of the centroid calculation, we need to calibrate accurately the influence function between slit-viewer offsets and field-selector offsets. For this purpose, we record several images of a calibration source with the slit-viewer, in closed-loop and with the source image off the slit by 0.5", then with an offset of 1" in different directions.

The matrix SV-centroid offset vs. field selector offset is then inverted, and is used as the SV-FS control matrix of the Figure 6.

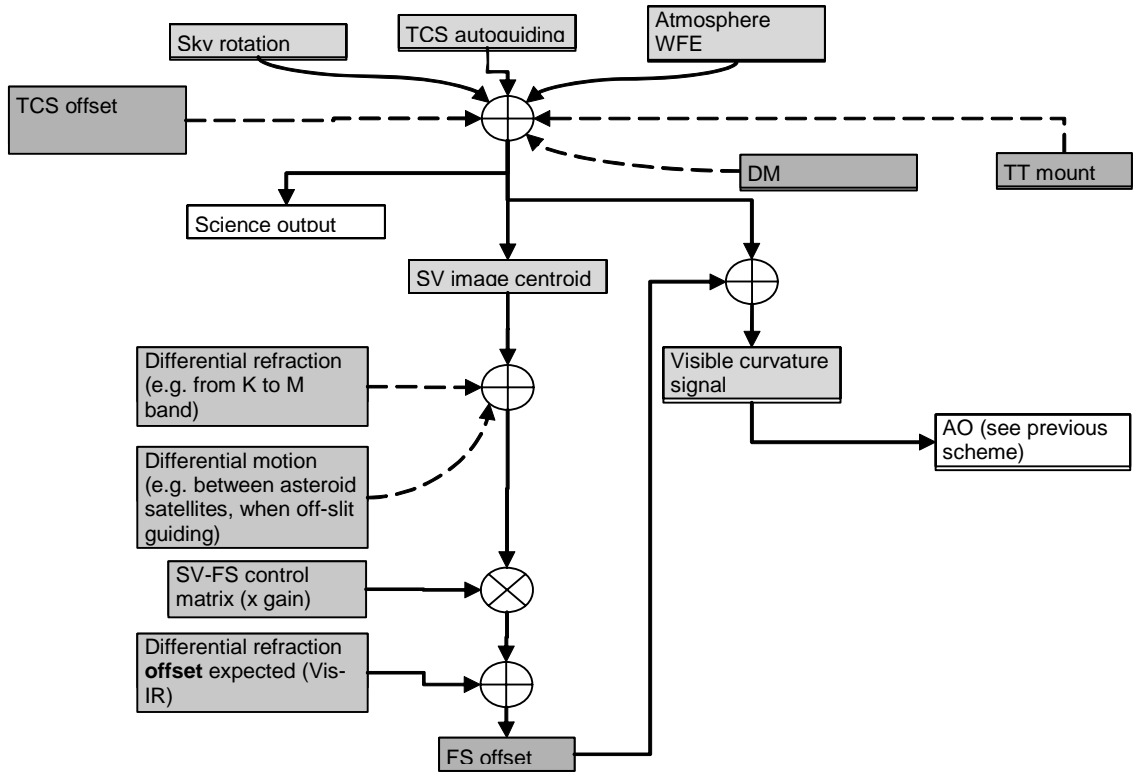

**Figure 6: slit-viewer / field selector loops**

## 5. Commissioning process, ITC

During the commissioning of the warm part of CRIRES (including the AO), we needed a test camera, with a sampling allowing us to resolve the diffraction spot. We therefore used the ITC (infrared test camera) that our group used for the commissioning of all the MACAO-systems, providing in our case a pixel scale of approximately 17 mas/pixel. The camera was used in its f/15 configuration, and the camera's Strehl ratio ranges between ~80% and ~90%, depending on the area used on the detector. In the following section 6, the results presented in terms of Strehl ratio are not corrected from the Strehl of the camera, unless otherwise specified. The results presented in EqE are not corrected for this camera effect, and their influence is anyway marginal in most cases.

Besides, in the case of CRIRES, we modified slightly the instrument-SW configuration parameters, so that we were able to offset the telescope interactively, the same way CRIRES will do on-sky. Of course, this was requiring having the derotator properly configured.

Briefly summarized, the installation of the instrument on the platform consisted in different main parts:
- installation of the main structure, the breadboard, the optical components, the cooling and power lines, electronics rack, …
- alignment of the breadboard, then of the derotator with respect to the telescope axis,
- alignment of the warm optics (relaying the Nasmyth focus to the entrance of the cryostat and of the WFS)
- alignment of the wavefront-sensor.
- In parallel to these activities, testing of all functions, installation and test of the instrument-SW, …

The alignment of the derotator had to be redone after the completion of these tasks (the pupil alignment had been overlooked), which made our short commissioning schedule very tight. Still, within three weeks, the instrument was completed, and only a few calibrations were missing, completed during the last days of the commissioning period. This tight schedule could not have been met, without the experience that the commissioning team acquired on the MACAO-systems previously.

## 6. Results obtained

During the 6 nights used in this commissioning period, a comprehensive set of data has been acquired, providing a good assessment of the performance of the system. The Strehl ratio and performance achieved with MACAO-CRIRES are summarized in the Table 1. Note that rather high values of EqE (ensquared energy) are measured even in open-loop. This is again an indication that the outer scale effect plays a critical role in infrared image formation[viii], reducing drastically the FWHM of open-loop images.

| Result with respect to the R-magnitude | R = 10 | R = 13.5 | R = 15 | R = 17.4 | Open-loop |
|---|---|---|---|---|---|
| Median EqE, 0.2" side box, good seeing | 63% | 58% | (-) | (-) | 27% |
| Median EqE, 0.2" side box, bad seeing | 56% | 50% | 43% | 30% | 15% |
| Best Strehl ratio measured on the ITC images | 61% | 47% | 25% | (6%) | (-) |

**Table 1: performance summary. When no data were available, (-) is used; low Strehl ratio (<10%) are qualitative only, and in brackets. Bad seeing corresponds to a DIMM-seeing >0.73" (up to 1.2"), while good seeing conditions are for a DIMM-seeing<0.78". $\tau_0$ was not considered in this table, but was typically ranging between 2 and 4 ms during the commissioning period.**

The AO-performance is often referred to as a Strehl function of the guide-star magnitude. This plot is given in the Figure 7, where the Strehl values obtained for bright stars show a performance at the level of all other MACAO systems. Unfortunately, during the commissioning, the faint star observations were performed only with bad seeing conditions, and explain the fall of the performance seen for R>14. Complementary data will be taken to confirm that the EqE is as the level expected, during the following commissioning.

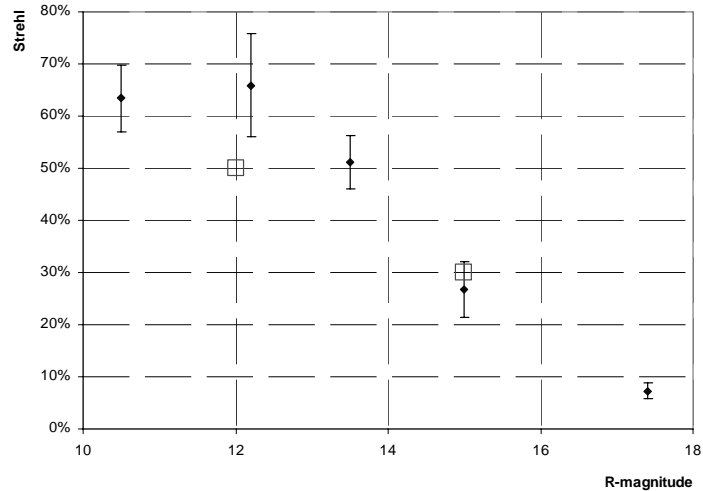

**Figure 7: Strehl ratio vs magnitude, as estimated from commissioning data. The faint star values have been measured with bad seeing conditions, and are therefore slightly pessimistic. The Strehl values are here corrected from the aberrations of the ITC (which delivers a Strehl ratio of 83%): it is the Strehl delivered by the AO-system to the vessel.**

Of more relevance is the graph showing the relationship between EqE and Strehl ratio. A clear relationship can be established, providing a rule of thumb to evaluate the EqE provided, as a function of the correction provided by the system. There is a clear slope discontinuity when reaching 10 to 15% of Strehl ratio, above which the improvement in terms of EqE becomes more expensive in terms of SR (Strehl ratio) required. The curve provided shows as well a strong difference with respect to what should be expected from a simple model, where a coherent image with a fraction of energy of SR superimposes a gaussian halo with an FWHM at the uncorrected seeing level and with an intensity of (1-SR). We interpret this strong discrepancy by a combination of two factors:
- the SR computed on the images is raw, including the aberrations of the camera. These aberrations are mostly low-order, and are therefore very efficient at lowering the SR without affecting much the 0.2"-box EqE,
- the EqE model is computed for a Kolmogorov approach, and doesn't take into account any outer-scale effect, which has a major impact in K-band.

The result of the operation of an adaptive optics system can be assimilated to an increase of the outer scale in the resulting wavefront, which explains the importance of the second term. We will in the near future take into account this effect in the simulation which provides the EqE vs the Strehl, and compare with the results we got during the commissioning time.

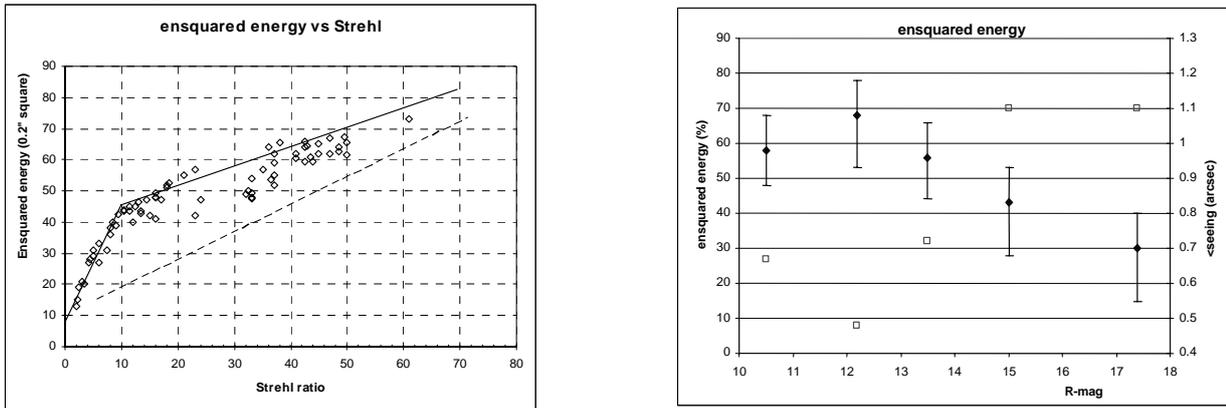

**Figure 8, left: ensquared energy as a function of the Stehl ratio provided by the system. Typical values for open-loop measurements provided 15 to 25% of EqE.**
**Right: ensquared energy versus R-magnitude. The averaged r0 value for the measurements done at each magnitude are drawn with squares, while the diamonds represent the ensquared energy curve.**

Besides this performance assessment, some targets have been observed for checking the performance of correction on extended sources, as well as on double stars, which results are given in the following sections.

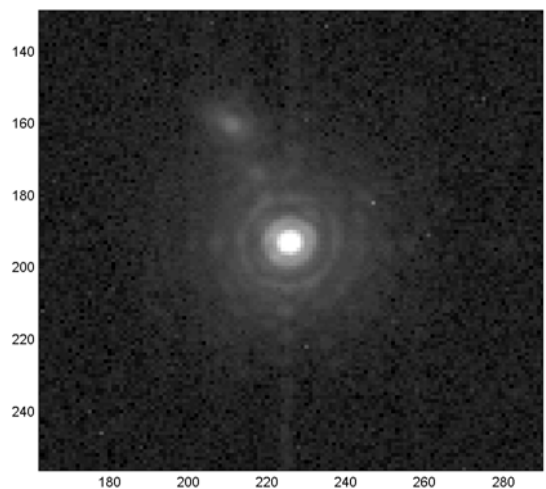

**Figure 9: the best star image obtained by MACAO-CRIRES; the star is a R-mag 12 star, and shows a Strehl ratio of 61%. Corrected from the aberrations introduced by the ITC, the Strehl provided by the AO reaches of at least 67%. The spot on the upper left is a ghost due to reflections in the K-narrow filter of the camera. The light scale is not linear, for emphasizing the visiblity of the Airy rings. The total integration time is around 15 s, and the image is only dark-subtracted.**

## 6.1. Double stars

In AO, the problem of double star observation is recurrent. For MACAO-CRIRES, we have explored the conditions for stability when a double star is observed. The main conclusions are the following:
-   a double star with a separation noticeably smaller than the field of view will produce stable images
-   a double star with a separation between 0.5 and 0.8 times the field of view will show a performance degradation with respect to the single star case, but will be stable,
-   a double star with a separation between 0.8 and ~2 times the field of view will show an unpredictable performance-stability, with the possibility of creation of ghost images of the double star, creating a triple star or worse
-   a double star with a separation above 3 times the field of view will be stably tracking on one of the two stars.

Those values were acquired on stars with similar brightness. Of course, this effect will be strongly mitigated by an important magnitude difference, but can be used as a guideline to try and improve the image quality in case a double star or an extended object are observed, by modifying the diaphragm diameter.

## 6.2. Galilean satellites of Jupiter

To complete the solar system views given by the MACAO systems during the previous commissioning times, we took some snapshots of the Galilean satellites of Jupiter, and the results were quite surprising, showing plenty of details where no surprise was expected, and revealing a new volcano on Io's surface, competing in brightness with Io's surface in K-band.

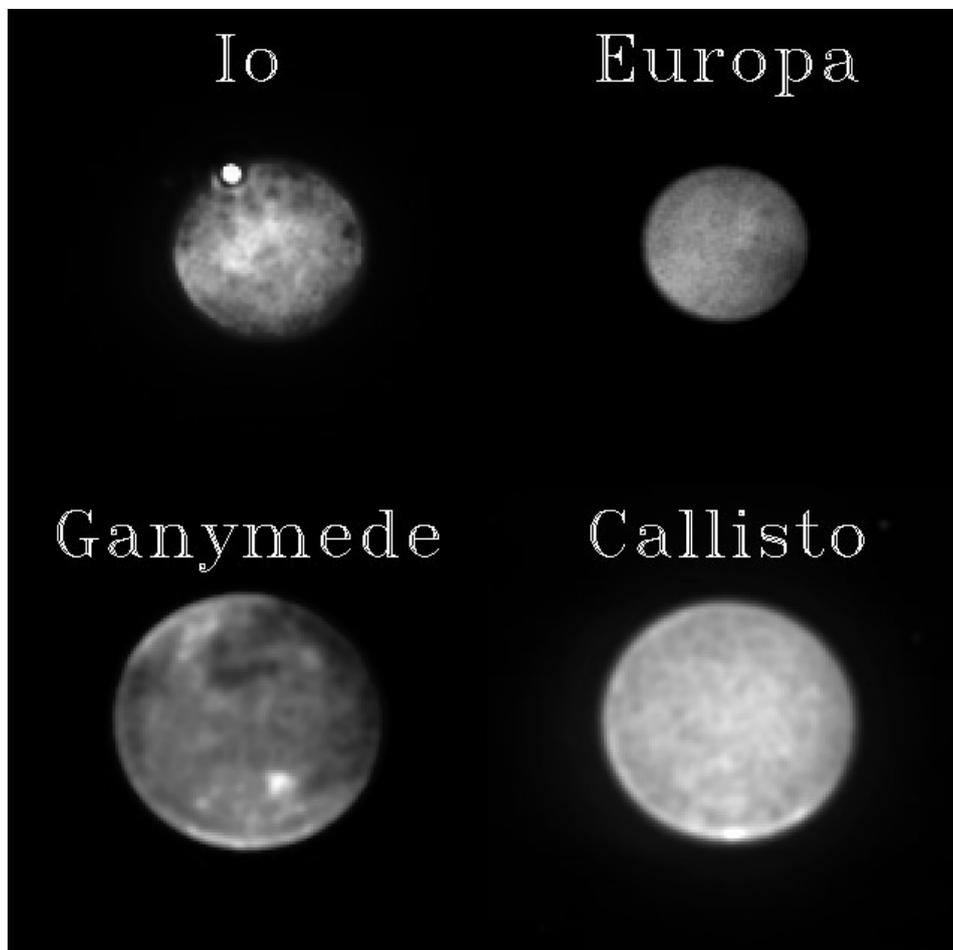

**Figure 10: Galilean satellites of Jupiter. The volcano seen on Io has not been identified as one of the usually active of Io. The diameter of Io was 1.1" at the time of the observation, Ganymede was 1.55". All satellites have the same scale in this image.**

### 6.3. Frosty Leo

Frosty Leo is a planetary nebula, around a post-AGB star, and it is a target which has already been caught during one of the MACAO-VLTI commissioning. A few years later, our new snapshot of this nebula shows that the performance of the MACAO systems has been slightly improved on some difficult targets.

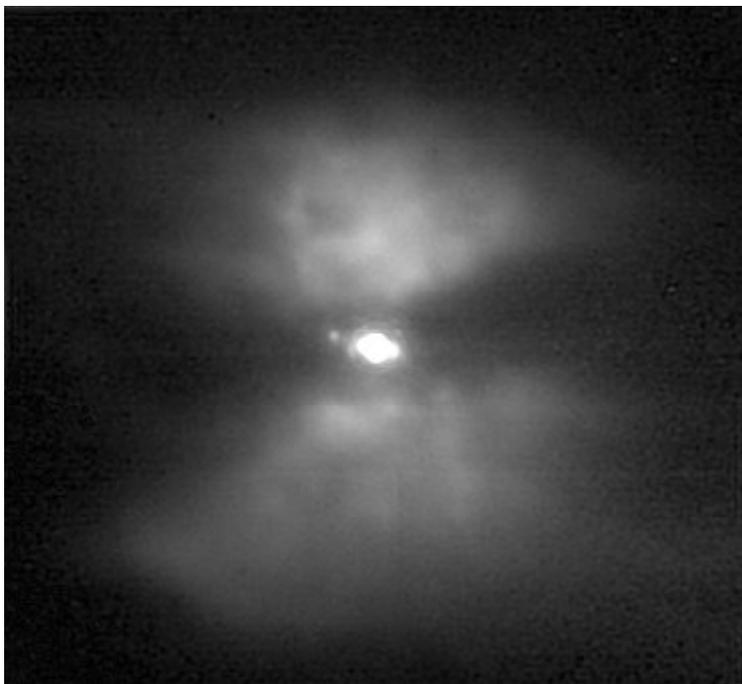

**Figure 11: Frosty-Leo. The FWHM of the star is <70 mas, in spite of the difficulty of the target.**

### 7. Next steps for CRIRES, conclusions

At the time of writing, the cold part of CRIRES has joined the warm part on the Nasmyth platform of UT1 in Paranal, and the commissioning observations with the whole instrument will start, with the aim to start the science operation in 2007. By that time, the AO will bring a two-folded gain in throughput for star-observations.
Concerning the performances of the system, the last of the MACAO systems has shown performances which are very consistent with respect to the previous MACAO systems in operation. We should as well note that the increased throughput provided by the adaptive optics will make possible a gain of more than one magnitude on the limiting magnitude for CRIRES, allowing catching up to K=17 stars at an unprecedented spectral resolution.

### 8. Aknowledgements

All our thanks are directed to the people who allowed this system to be brought in such a state. Both people from the AO group of ESO who opened the way to MACAO-CRIRES on the VLT, and the team in Paranal who was very supporting. Taking the risk of forgetting some names, here is a short list, in random order:
S. Stroebele, S. Guisard, Ü. Weilenmann, E. Bendek, N. Haddad, P. Amico, A. Smette.

---